\documentstyle[epsfig]{aipproc}

\begin{document}
\title{Dynamics of Cosmological Models\\ in the Brane-world Scenario}

\author{Antonio Campos$^*$ and Carlos F. Sopuerta$^*$}
\address{$^*$Relativity and Cosmology Group, \\
Portsmouth University, Portsmouth PO1 2EG, Britain}

\maketitle

\begin{abstract}
We present the results of a systematic investigation of the qualitative 
behaviour of the Friedmann-Lema{\^\i}tre-Robertson-Walker (FLRW) and 
Bianchi I and V cosmological models in Randall-Sundrum brane-world type 
scenarios.
\end{abstract}

Recently, Randall and Sundrum have shown that for non-factorizable 
geometries in five dimensions the zero-mode of the Kaluza-Klein 
dimensional reduction can be localized in a four-dimensional submanifold 
\cite{RanSun:1999b}.
The picture of this scenario is a five-dimensional space 
with an embedded three-brane where matter is confined and Newtonian 
gravity is effectively reproduced at large distances.

Here, we summarize the qualitative behaviour of FLRW and
Bianchi I and V cosmological models in this scenario
(see \cite{CamSop:2001} for more details).
In particular, we have studied how the dynamics changes with respect 
to the general-relativistic case. 
For this purpose we have used the formulation introduced 
in~\cite{ShiMaeSas:2000}.
{}From the Gauss-Codazzi relations the
Einstein equations on the brane are modified with two
additional terms.
The first term is quadratic in the matter variables and the second
one is the electric part of the five-dimensional Weyl tensor.
In this communication we will  consider the effects due to the
first term.
The study including both corrections has been carried out 
in~\cite{CamSop:2001b}.
We also assume that the matter content is described
by a perfect fluid with energy density, $\rho$, and pressure, $p$,
related by a linear barotropic equation of state,
$p=(\gamma - 1)\rho$ with $\gamma \in [0,2]$.

\begin{figure}[t!]
\centerline{\epsfig{file=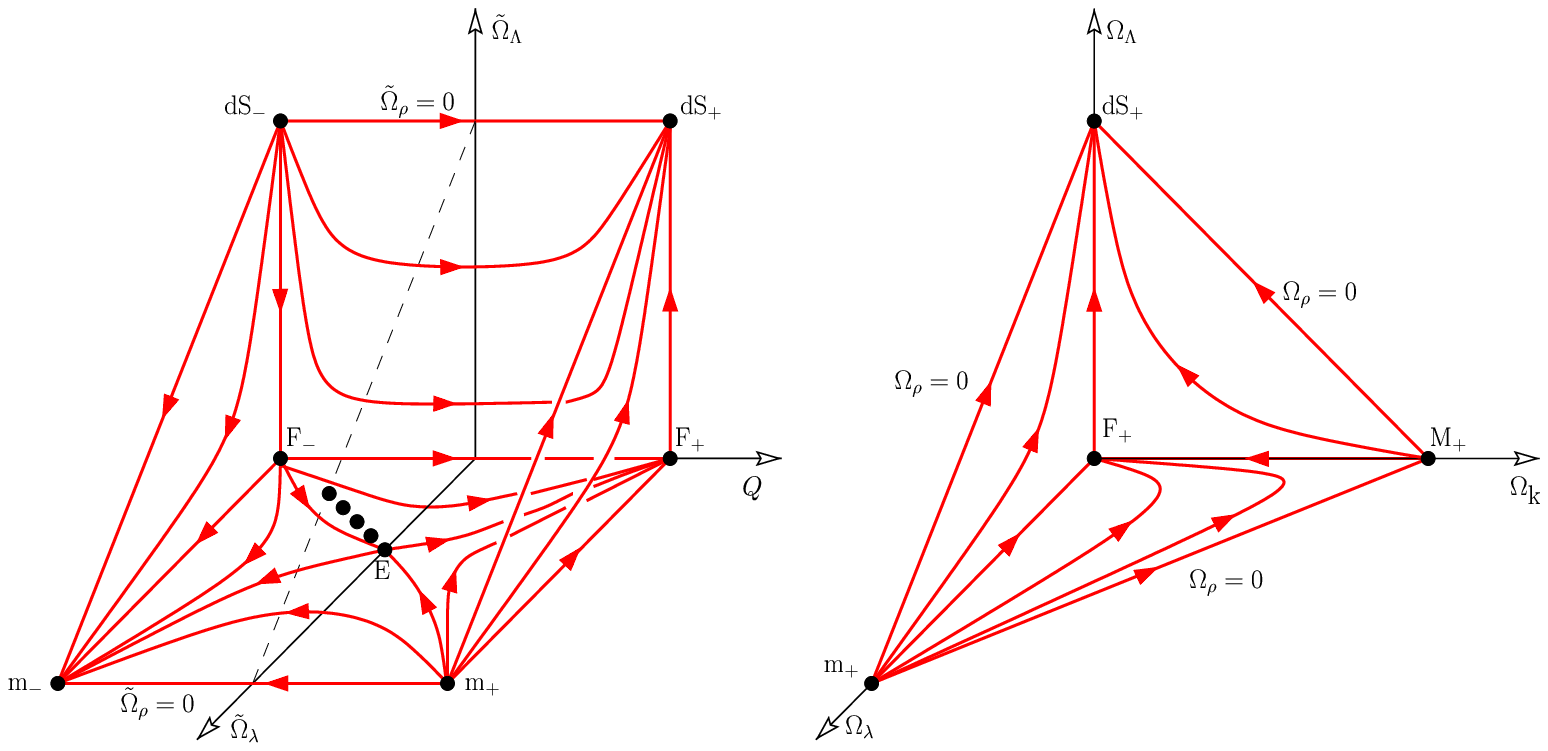,height=2.5in,width=4.75in}}
\vspace{10pt}
\caption{State space for the FLRW models with 
$\gamma\in (\frac{1}{3},\frac{2}{3})$ and non-negative 
(left) and non-positive (right) spatial curvature.
The variables $(\Omega_\rho, \Omega_k, \Omega_\Lambda, \Omega_\lambda)$,
and their analogs with tilde, are fractional contributions of the
energy density, spatial curvature, cosmological constant and brane
tension, respectively, to the universe expansion~[2]. 
Replacing $\Omega_k$ by $\Omega_\sigma$ (the shear contribution) and 
$\mbox{M}_+$ by $\mbox{K}_+$, the drawing on the right is also the state 
space for Bianchi I models with $\gamma\in (1,2)$.
For clarity, only trajectories on the invariant planes have been drawn.
The dynamics of a general trajectory can be inferred from them.
The subscript ``$+$'' (``$-$'') refers to the expanding (contracting)
character of the models.
The planes $\tilde{\Omega}_\lambda = \Omega_\lambda = 0$ 
correspond to the state space of general relativity.} 
\label{esp3}
\end{figure}

When the brane dynamics is described by a FLRW model
we find five generic critical points:
the flat FLRW models ($\mbox{F}$);
the Milne universe ($\mbox{M}$); 
the de Sitter model ($\mbox{dS}$);
the Einstein universe ($\mbox{E}$);
and the non-general-relativistic 
Bin$\acute{\mbox{e}}$truy-Deffayet-Langlois
(BDL) model ($\mbox{m}$)~\cite{BinDefLan:2000}.
The dynamical character of these critical points and the
structure of the state space depend on the equation
of state, or in other words, on the parameter $\gamma$.
This means that we have bifurcations for some values of
$\gamma$, namely $\gamma=0, \textstyle{1\over3}, \textstyle{2\over3}$.
The bifurcation at $\gamma=\textstyle{1\over3}\,$ is a genuine feature
of the brane world and is characterized by the appearance of an infinite 
number of non-general-relativistic critical points.
The Einstein Universe critical point appears for 
$\gamma\geq \textstyle{1\over3}$, in contrast with
the general-relativistic case, where it appears for 
$\gamma\geq\textstyle{2\over3}$.
Actually, for $\textstyle{1\over3} < \gamma < 2$ we 
do not have an isolated critical point corresponding to the Einstein
universe but a line of critical points, as can be seen in the state 
space shown in Figure~\ref{esp3}.
Another important feature of these scenarios is that the 
dynamical character of some of the points changes.  
For instance, the expanding and contracting flat FLRW models, 
which in general relativity are repeller and attractor for 
$\gamma>\textstyle{2\over3}$, 
are now saddle points for all values of $\gamma$.
The new non-general-relativistic critical point,
the BDL solution~\cite{BinDefLan:2000}, 
describes the dynamics near the initial Big-Bang singularity and,
for recollapsing models, near the Big-Crunch singularity.
More precisely, the dynamical behaviour near these singularities
is governed by a scale factor $a(t)=t^{1/(3\gamma)}$ which differs
from the standard evolution in general-relativistic cosmology,
where $a(t)=t^{2/(3\gamma)}$.
Finally, the general attractor for ever expanding universes is,
as in general relativity, the de Sitter model. 
For recollapsing
universes, which now appear for $\gamma>\textstyle{1\over3}$,
the contracting BDL model is the general attractor.
However, if we only consider the invariant manifold representing
general relativity, the contracting Friedmann universe is the
general attractor for $\gamma > 2/3$.
On the other hand,
for zero cosmological constant and $\gamma < 2/3$ the expanding 
Friedmann universe is also an attractor.

For the homogeneous but anisotropic Bianchi I and V 
cosmological models, which contain the flat and negatively 
curved FLRW models respectively, we find an additional critical 
point, namely the Kasner vacuum spacetimes ($\mbox{K}$).
In the Bianchi I case the state space can be represented by the 
same type of drawings used for the non-positive spatial
curvature sector of the FLRW evolution (see Figure~\ref{esp3}).
A representative set of diagrams for Bianchi V models is given 
in~\cite{CamSop:2001}.
For Bianchi I models we have found a new bifurcation at $\gamma=1$
and for Bianchi V models at $\gamma=\textstyle{1\over3},1\,$,
in addition to the general relativity bifurcations at
$\gamma=0,2\,$ and $\gamma=0,\textstyle{2\over3},2\,$, 
respectively.
Some of the dynamical features explained above for the FLRW are 
shared by the these Bianchi models.
However, the most interesting point here
is the possibility of studying the dynamics of anisotropy in
brane-world scenarios.
Specifically, we have seen~\cite{CamSop:2001} that, although 
now we can have intermediate stages in which the anisotropy grows,
expanding models isotropize as it happens in general relativity. 
This is expected since the energy density decreases and hence, the
effect of the extra dimension becomes less and less important.
The situation near the Big Bang is more interesting.  
In the brane-world scenario
anisotropy dominates only for $\gamma<1$, whereas in general
relativity dominates for all the physically relevant values
of $\gamma$.

To conclude, let us summarize the main features of the dynamics of 
cosmological models on the brane. 
First, we have found
new equilibrium points, the BDL models~\cite{BinDefLan:2000}, representing 
the dynamics at very high energies, where the extra-dimension 
effects become dominant.
Thus, we expect them to be a generic feature of the 
state space of more general cosmological models in the 
brane-world scenario. 
Second, the state space presents new bifurcations for some particular
equations of state.
Third, the dynamical character of some of the critical points
changes with respect to the general-relativistic case.
Finally, for models in the range $1< \gamma \leq 2$, that is for models 
satisfying all the ordinary energy conditions and causality requirements, 
we have seen that the anisotropy is negligible near the initial 
singularity. 
This naturally leads to the questions of whether the oscillatory
behaviour approaching the Big Bang predicted by general relativity
is still valid in brane-world scenarios.
We are currently investigating this issue by considering Bianchi IX
cosmological models~\cite{CamSop:2001c}.

\vskip5mm\noindent
{\bf Acknowledgments:}
This work has been supported by the European Commission (contracts 
HPMF-CT-1999-00149 and HPMF-CT-1999-00158).

\end{document}